\documentclass[conference]{IEEEtran}

\IEEEoverridecommandlockouts

\usepackage{amsmath,amssymb,amsfonts}
\usepackage{algorithmic}

\usepackage{booktabs}

\usepackage{cite}
\usepackage{comment}

\usepackage{graphicx}
\usepackage{textcomp}
\usepackage{xcolor}
\usepackage{xspace}

\def \Budget       {\kappa_{\TimeIdx}}
\def \BWD          {b}

\def \CAP       {\textsc{cap}}
\def \ConfigSet {P}
\def \ConfigSetSolution {\Gamma}
\def \ConfigIdx {\gamma}
\def \Capl      {\textsc{cap}_{\ell}}

\def \DemandIdx     {d}
\def \DemandSet     {D}
\def \DST           {\textsc{dst}}
\def \DualIdx       {u}

\def \InterventionIdx   {Q_\TimeIdx}

\def \LinkSet   {L}
\def \LinkSett  {L_{\TimeIdx}}

\def \LinkIdx   {\ell}

\def \LambdaL   {\lambda_{\LinkIdx}}

\def \LinkWeight    {\weight_\LinkIdx}

\def \maxseg {\texttt{maxseg}\>}

\def \NameMILP      {SR\_CGP} 

\def \SRC           {\textsc{src}}
\def \SegmentKnIdx  {\sigma}

\def \SegmentKtSet  {L^{K_{\TimeIdx}}}

\def \TimeIdx       {t}

\def \weight       {\textsc{w}}

\def \zLPOPTt       {z_{\textsc{lp}}^{\star \> t}}
\def \zILPUBt       {z_{\textsc{ilp}}^{\textsc{ub} \> t}}
\def \zILPOPTt      {z_{\textsc{ilp}}^{\star \> t}}
\def \zLB           {z_{\textsc{lb}}}
\def \zbest         {z_{\text{best}}}
\def \zLBt           {z_{\textsc{lb}}^t}
\def \zbestt         {z_{\text{best}}^t}

\newcommand{\update}[1]{{\color{pink} #1}}

\newcommand{\BJ}[1]{\noindent{\color{red}[\textsc{BJ:} #1]}\xspace}
\newcommand{\PT}[1]{\noindent{\color{blue}[\textsc{Phill:} #1]}\xspace}
\newcommand{\Junior}[1]{\noindent{\color{cyan}[\textsc{Junior:} #1]}\xspace}

\def\BibTeX{{\rm B\kern-.05em{\sc i\kern-.025em b}\kern-.08em
T\kern-.1667em\lower.7ex\hbox{E}\kern-.125emX}}
\begin{document}

\title{Segment Routing Traffic Engineering with Time-Based Reconfiguration Constraints}
\author{\IEEEauthorblockN{Brigitte Jaumard, Nguyen Phuc Tran and Junior Momo Ziazet}
  \IEEEauthorblockA{\textit{Computer Science and Software Engineering} \\
    \textit{Concordia University}\\
    Montreal (QC) Canada \\
  brigitte.jaumard@concordia.ca}
}

\maketitle

\begin{abstract}
Segment routing (SR) involves routing requests through shortest paths or a limited number of segments, which are themselves the shortest paths between their endpoints. While this flexibility makes segment routing attractive for traffic engineering (TE) in IP/MPLS networks, it becomes challenging to update when some links become unavailable (e.g., maintenance) and routing decisions must adapt while limiting the number of configuration changes between two consecutive periods.

We address the problem of dynamically selecting SR-TE configurations for multiple traffic demands with the objective of minimizing the maximum link utilization (MLU).
In addition, a reconfiguration budget is imposed in order to limit the number of segments modified.

Based on this representation, we develop a MLU column-generation optimization framework.
It allows to jointly consider traffic distribution, SR configuration complexity, and temporal reconfiguration costs in time-varying IP/MPLS networks.

The numerical results are obtained using realistic datasets provided by Orange, with topologies containing up to 1,263 nodes and 15,000 traffic demands.
The optimality gap (i.e., accuracy) is below 5\% in 80.45\% of the 133 instance-time evaluations.

\end{abstract}


\section{Introduction}

Traffic engineering (TE) plays a central role in the operation of Internet Protocol (IP) networks by determining how traffic demands are routed over available network resources. 
As current networks are evolving towards service-driven networks, 
Segment routing (SR) is now widely used because it is recognized as a TE technology that allows for significant simplifications and better adaptability \cite{Filfils_2013, Anand_2017}. 
Differently from classical MultiProtocol Label Switching (MPLS) routing, segment routing (SR) divides a forwarding path into segments, each encoded as an ordered list in the packet header. 
This approach ensures that the network only needs to maintain minimal state.

Numerous studies have been devoted to segment routing, as well as its protection, restoration, and adaptability under various conditions.
Indeed, as traffic volumes and network conditions vary over time, operators must continuously adapt routing decisions to avoid congestion and maintain efficient utilization of network capacity. Often, there is a need for some interventions, to repair a link failure, or to modify the capacity of some links/nodes. 
The segment routing then transforms into a problem of timely segment reconfiguration, for a small number of segments at a time, with the fundamental objective of distributing traffic across the network in order to avoid heavily congested links, which leads to the classic objective of minimizing the highest link load.

We propose a column generation framework with a decomposition over the set of demands, and with some budget constraints when network repair or update are required during a given time window. Routing of each demand is handled efficiently in each pricing problem on a segment graph made of precomputed shortest paths in order to ease the segment routing calculations and avoid multiple recalculations. Reconfiguration budget is handled as a single linear constraint in spite of its nonlinear definition.

The paper is organized as follows. Next section reports the most significant recent work. 
Section \ref{sec:PbStatement} provides the formal problem statement.
The proposed mathematical model and its solution are described in Section \ref{sec:formulation}.
Numerical results are reported in Section \ref{sec:results}.
Conclusions are drawn in the last section.


\section{Literature Review}

Different approaches have been proposed for solving the segment routing design problem, from ILP and MILP formulations including Dantzig-Wolfe decomposition \cite{Jadin_2019}, heuristics and meta-heuristics \cite{Moreno_2017}, as well as AI/ML approaches \cite{Tian_2025}. 

Regardless of the objective chosen, ILP and MILP formulations quickly encounter scalability problems, which is why most mathematical programming studies consider decomposition techniques as an alternative for solving larger data instances. Decomposition is made with respect to demands, and consequently the subproblem is a shortest path problem governed by its objective, the so-called reduced cost (see, e.g., Chavtal \cite{chv83} if not familiar with decomposition techniques). The differences lie in the objective (MLU for \cite{Li_2020}, throughput maximization for \cite{Jadin_2019}).

Heuristics are usually defined in order to speed up the ILP/MILP formulations, or as a stand alone, see, e.g., Moreno \textit{et al.} \cite{Moreno_2017}, and the recent survey of Wu \textit{et al.} \cite{Wu_2023}.

Additionally, some studies leverage machine learning (ML) techniques, especially deep reinforcement learning (DRL), for accelerated problem-solving by learning to make TE decisions directly, see, e.g., \cite{Li_2024, Pei_2024, Tian_2025}. 
The most recent paper, Tian \textit{et al.} \cite{Tian_2025} consider relatively small network topologies ($\leq 23$ nodes) on which DRL remains at an average distance of 8\% to 34\% from the optimal solution of a 2-SR-ILP model. 

While few papers studied resilient segment routing, e.g., \cite{Schuller_2021}, there is no study on the planning of interventions, i.e., scheduled maintenance operations, the focus of this study and of the 2026 Orange competition~\footnote{https://roadef.org/challenge/2026}. 


\section{Problem Statement}
\label{sec:PbStatement}

We consider a time-varying traffic engineering problem in an IP/MPLS network modeled as a directed graph
$G=(V,\LinkSet)$, where $V$ denotes the set of network nodes and $\LinkSet$ denotes the set of directed links. Each link $\LinkIdx \in \LinkSet$ is characterized by a transport capacity $\Capl$ and a routing weight $\LinkWeight$, which is used by the routing protocol to compute shortest paths.

At a given time $\TimeIdx$, a subset of links $\InterventionIdx \subseteq \LinkSet$ may become unavailable due to maintenance, failures, or other operational events, collectively referred to as \emph{interventions}. The resulting network is therefore represented by the time-dependent graph
\begin{equation}
G_{\TimeIdx}=(V,\LinkSett),
\qquad
\LinkSett=\LinkSet\setminus\InterventionIdx.
\end{equation}

A set of traffic demands $\DemandSet_{\TimeIdx}$ must be routed over $G_{\TimeIdx}$. Each demand $\DemandIdx\in\DemandSet_{\TimeIdx}$ is characterized by a source node $\SRC_\DemandIdx\in V$, a destination node $\DST_\DemandIdx\in V$, and a traffic volume $\BWD_\DemandIdx$ (Gb/s), i.e.,
$\DemandIdx =
(\SRC_\DemandIdx,\DST_\DemandIdx,\BWD_\DemandIdx)$.

\begin{figure}[htb]
    \centering
    \includegraphics[width=0.8\linewidth]{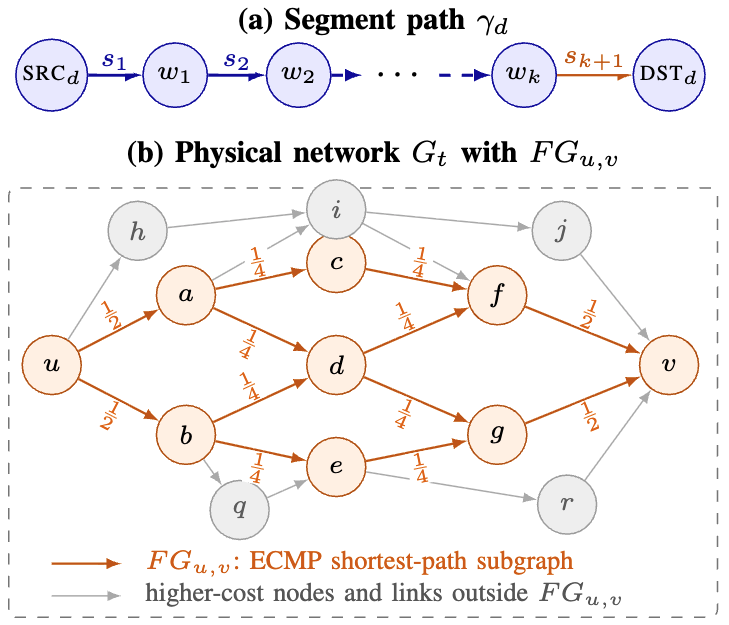}
    \caption{An example of Segment-routing configuration and ECMP forwarding graph.}
    \label{fig:FowardingGraph}
\end{figure}

To route a demand $\DemandIdx$, we use a segment-routing configuration represented by a segment path
$\ConfigIdx_{\DemandIdx}$ =
$(\SRC_\DemandIdx,w_1,\ldots,w_k,\DST_\DemandIdx)$,
where $w_1,\ldots,w_k$ are intermediate waypoints, and $k=0$ corresponds to a path without explicitly specified waypoints. Each pair of consecutive nodes $(u,v)$ in $\ConfigIdx_{\DemandIdx}$ defines a \emph{segment}, as illustrated in Figure~\ref{fig:FowardingGraph}(a). 
The forwarding behavior associated with a segment $(u,v)$ is represented by a forwarding graph $FG_{u,v}$, which is the subgraph of $G_{\TimeIdx}$ induced by all links belonging to shortest paths from $u$ to $v$, including paths resulting from Equal-Cost Multi-Path (ECMP) routing. See Figure \ref{fig:FowardingGraph}(b) for an illustration.

The forwarding graphs of the segments that define $\ConfigIdx_{\DemandIdx}$ determine the set of physical links used by the traffic of $\DemandIdx$. Therefore, the traffic carried by each link depends on the segment paths selected for the set of demands.

At each time $\TimeIdx$, the primary objective is to minimize the \emph{maximum link utilization} (MLU) in the available network $G_{\TimeIdx}$. Let the utilization of link $\LinkIdx\in\LinkSett$ be denoted by $\LambdaL$. The MLU is defined as
\begin{equation}
\mathrm{MLU}_{\TimeIdx}=
\max_{\LinkIdx\in\LinkSett}
\LambdaL.
\end{equation}

The optimization is subject to two operational constraints. First, routing changes must satisfy a strict reconfiguration budget. Specifically, between two consecutive routing configurations at times $\TimeIdx-1$ and $\TimeIdx$, the total number of waypoint additions and removals must not exceed a predefined budget $\Budget$. This constraint limits the amount of routing reconfiguration required to adapt to network changes. Second, the complexity of each segment path is bounded: for every demand $\DemandIdx\in\DemandSet_{\TimeIdx}$, the number of segments in $\ConfigIdx_{\DemandIdx}$ must not exceed a predefined maximum $\maxseg$.

Therefore, at each time $\TimeIdx$, the problem consists of selecting a segment path $\ConfigIdx_{\DemandIdx}$ for every demand $\DemandIdx\in\DemandSet_{\TimeIdx}$ such that the resulting routing configuration minimizes the MLU while satisfying the reconfiguration budget and the maximum number of segments per demand. Formally,  given $\ConfigSetSolution_{\DemandSet_{\TimeIdx-1}}$, the segment-routing
configurations for all demands at $\TimeIdx-1$, the problem can be stated as: 
\begin{alignat}{4}
    \min\limits_{\ConfigSetSolution_{\DemandSet_{\TimeIdx}}} \quad 
&   \mathrm{MLU}_{\TimeIdx} \\ 
\text{s.t.} \quad 
& \mathrm{Reconf}
\left(
\ConfigSetSolution_{\DemandSet_{\TimeIdx-1}},
\ConfigSetSolution_{\DemandSet_{\TimeIdx}}
\right)
\leq \Budget \quad \\
&   |\ConfigIdx_{\DemandIdx}| - 1 \leq \maxseg
&&   \DemandIdx \in \DemandSet_{\TimeIdx} \\
&   \ConfigIdx_{\DemandIdx} \text{ is feasible in } G_{\TimeIdx} 
&&   \DemandIdx \in \DemandSet_{\TimeIdx}.
\label{eq:pb}
\end{alignat}
Here, $\mathrm{Reconf}(\cdot,\cdot)$ denotes the number of segment additions and removals required to transform the routing configuration at time $\TimeIdx-1$ into the configuration selected at time $\TimeIdx$. The feasibility condition ensures that every selected segment and its associated forwarding graph are defined over the available network $G_{\TimeIdx}$.


\section{Proposed Mathematical Model and Solution}
\label{sec:formulation}

The key idea of the proposed mathematical model is to transform the segment-routing
problem into a path-selection problem on an auxiliary graph.
Such a model allows the use of a column-generation algorithm in which
each potential segment-routing configuration for a given demand is represented by a column and an associated decision variable.


\subsection{Segment Graph}

For a demand $\DemandIdx$, each pair of consecutive nodes $(u,v)$ in a
segment-routing configuration $\ConfigIdx_{\DemandIdx}$ defines a segment
from $u$ to $v$. Under ECMP, this segment may correspond to multiple physical
paths in the available network $G_{\TimeIdx}$. We therefore construct an
auxiliary \emph{segment graph}
\begin{equation}
K_{\TimeIdx} = (V,\SegmentKtSet),
\end{equation}
whose node set is identical to that of the physical network
$G_{\TimeIdx}$. Each arc $\SegmentKnIdx \in \SegmentKtSet$ represents a
feasible segment between two distinct nodes $(u,v)$, i.e.,
\begin{equation}
\SegmentKnIdx = (u,v),
\qquad
u,v \in V,\quad u \neq v.
\end{equation}

Associated with each segment $\SegmentKnIdx=(u,v)$ is a forwarding graph
$FG_{u,v}$, which captures the set of physical links used by the ECMP
shortest paths from $u$ to $v$. Consequently, a path in $K_{\TimeIdx}$
corresponds directly to a segment-routing configuration in
$G_{\TimeIdx}$. In particular, for a demand
\begin{equation}
\DemandIdx =
(\SRC_{\DemandIdx},\DST_{\DemandIdx},\BWD_{\DemandIdx}),
\end{equation}
finding a feasible segment-routing configuration is equivalent to finding
a path from $\SRC_{\DemandIdx}$ to $\DST_{\DemandIdx}$ in
$K_{\TimeIdx}$. The selected segments can then be mapped to the physical
links of $G_{\TimeIdx}$ through their corresponding forwarding graphs.

\subsection{Column-Generation Framework}

The number of feasible segment-routing configurations can be very large,
as every path between the source and destination in $K_{\TimeIdx}$ defines a
potential configuration. Enumerating all configurations for all demands
would therefore result in a prohibitively large optimization problem.

To address this issue, we propose a decomposition model, referred to as
\emph{\NameMILP}, based on column generation. In this framework, a
\emph{configuration} represents one feasible segment-routing configuration
for a demand and is treated as a column in the master problem.

Let
$\ConfigSet_{\DemandIdx,\TimeIdx}$ denote the set of configurations currently
available for demand $\DemandIdx$ at time $\TimeIdx$. Rather than
enumerating the complete configuration space, the column-generation
procedure starts with a restricted set of configurations and iteratively
expands it until an optimality condition is satisfied.

The framework consists of two interacting components:

\begin{itemize}
    \item The \emph{master problem} selects exactly one configuration for
    each demand from the current configuration sets and determines the
    routing configuration that minimizes the MLU while satisfying the
    network-capacity and reconfiguration constraints.

    \item The \emph{pricing problem} searches for additional feasible
    configurations with negative reduced cost. Such configurations have the
    potential to improve the current master solution and are added to the
    corresponding configuration set.
\end{itemize}

The master and pricing problems are solved alternately until no improving
configuration can be generated, (see Fig.~\ref{fig:cg_work_flow} for more details). 
Since each pricing problem can be solved
independently for each demand, the pricing problems can be solved in parallel.

\subsection{Master Problem: Configuration Selection}

For each demand $\DemandIdx \in \DemandSet_{\TimeIdx}$, an initial set of
feasible configurations is generated using shortest-path routing. These
configurations provide the initial columns of the restricted master
problem. Additional configurations are subsequently generated by the
pricing problem.

Given the current configuration sets
$\ConfigSet_{\DemandIdx,\TimeIdx}$, the master problem selects one
configuration for each demand.

\subsubsection{Parameters}

\begin{itemize}
    \item $\delta_{\LinkIdx\DemandIdx}^{\ConfigIdx}$: fraction of demand
    $\DemandIdx$ routed over physical link $\LinkIdx$ when configuration
    $\ConfigIdx$ is selected. These coefficients are computed during a
    preprocessing step by calculating the ECMP shortest paths between all
    relevant node pairs.

    \item $\varphi_{\SegmentKnIdx}^{\ConfigIdx}\in \{0,1\}$: indicates if configuration $\ConfigIdx$ contains segment $\SegmentKnIdx$.

    \item $\CAP_{\LinkIdx}$: transport capacity of physical link
    $\LinkIdx$.

    \item $\BWD_{\DemandIdx}$: traffic volume of demand $\DemandIdx$.

    \item $\kappa_{\TimeIdx}$: Maximum allowable routing reconfiguration, i.e., number of segment changes, between consecutive time periods.
\end{itemize}

\subsubsection{Decision Variables}

\begin{itemize}
    \item $y_{\DemandIdx}^{\ConfigIdx} \in \{0,1\}$: binary variable equal
    to $1$ if configuration $\ConfigIdx$ is selected for demand
    $\DemandIdx$, and $0$ otherwise.

    \item $\LambdaL \geq 0$: utilization ratio of
    physical link $\LinkIdx$. 
    

    \item $z \geq 0$: auxiliary variable representing the MLU.
\end{itemize}

\subsubsection{Mathematical Programming Model}

\begin{alignat}{2}
 & \min z && \nonumber \\
 & \LambdaL \leq z 
 && \LinkIdx \in \LinkSett \label{eq:max_utilization} \\
 & \sum_{\ConfigIdx \in \ConfigSet_{\DemandIdx,\TimeIdx}} y_{\DemandIdx}^{\ConfigIdx} = 1,
 && \DemandIdx \in \DemandSet_{\TimeIdx} \label{eq:onePathperDemand} \\
 & \sum_{\DemandIdx \in \DemandSet_{\TimeIdx}} \BWD_{\DemandIdx}
   \sum_{\ConfigIdx \in \ConfigSet_{\DemandIdx,\TimeIdx}} \delta_{\LinkIdx\DemandIdx}^{\ConfigIdx} y_{\DemandIdx}^{\ConfigIdx}
   \leq
   \LambdaL \CAP_{\LinkIdx} \quad
 && \LinkIdx \in \LinkSett \label{eq:linkCapa} \\
  &  \sum\limits_{\DemandIdx \in \DemandSet_{\TimeIdx}} 
     \sum\limits_{\ConfigIdx \in \ConfigSet_{\DemandIdx,\TimeIdx}}
     \sum\limits_{\SegmentKnIdx \in \SegmentKtSet}
     \left|
        \varphi_{\SegmentKnIdx}^{\ConfigIdx} 
        -
        \varphi_{\SegmentKnIdx}^{\Bar{\ConfigIdx}_{\DemandIdx, \TimeIdx-1}}
     \right|
     y_{\DemandIdx}^{\ConfigIdx}
     \leq \kappa_{\TimeIdx}
  && \label{eq:budget} \\
   & y_{\DemandIdx}^{\ConfigIdx} \in \{0,1\}
  && \hspace*{-1.cm} \DemandIdx \in \DemandSet_\TimeIdx, \ConfigIdx \in \ConfigSet \\
   & \lambda_{\LinkIdx}, z \geq 0 
  && \hspace*{-1.cm} \LinkIdx \in \LinkSet_\TimeIdx. \label{eq:defLambda}
 \end{alignat}



        

Constraints \eqref{eq:max_utilization} set the utilization MLU bounding of every link.
Constraints \eqref{eq:onePathperDemand} enforce the selection of exactly one configuration for every demand.
Link-capacity constraints \eqref{eq:linkCapa} state that the total traffic routed over each physical link cannot exceed its capacity. 
Lastly, constraints \eqref{eq:budget} express that the routing configuration is subject to a strict reconfiguration
budget (in terms of total change in segment usage across all demands)  between consecutive time periods. 

\subsection{Pricing Problem: Configuration Generation}

After solving the restricted master problem, the dual variables associated
with the configuration-selection, link-capacity, and reconfiguration
constraints are used to determine whether additional configurations can
improve the current solution.

For each demand $\DemandIdx$, the pricing problem searches for a feasible
path in the segment graph $K_{\TimeIdx}$ whose associated configuration has
negative reduced cost. Such a configuration is added as a new column to the
master problem. Since the pricing problem is independent for each demand,
the resulting pricing problems can be solved in parallel.

For readability, the demand index is omitted from the pricing formulation
below.

\subsubsection{Parameters}

\begin{itemize}
    \item $f_{\LinkIdx\SegmentKnIdx}$: fraction of traffic routed over
    physical link $\LinkIdx$ when segment $\SegmentKnIdx$ is selected.
    These coefficients are obtained during the preprocessing phase by
    computing the ECMP forwarding graphs for all feasible node pairs.

    \item $\DualIdx_{\DemandIdx}^{\eqref{eq:onePathperDemand}}$: dual
    variable associated with the configuration-selection constraint
    \eqref{eq:onePathperDemand}.

    \item $\DualIdx_{\LinkIdx}^{\eqref{eq:linkCapa}}$: dual variable
    associated with the link-capacity constraint \eqref{eq:linkCapa}.
    \item $\DualIdx_{\TimeIdx}^{\eqref{eq:budget}}$: dual variable
    associated with the reconfiguration budget constraint \eqref{eq:budget}.
\end{itemize}

\subsubsection{Decision Variables}

\begin{itemize}
    \item $\varphi_{\SegmentKnIdx} \in \{0,1\}$: binary variable equal to
    $1$ if segment $\SegmentKnIdx$ is selected in the generated
    configuration, and $0$ otherwise.


    \item $\delta_{\LinkIdx} \geq 0$: fraction of the demand routed over
    physical link $\LinkIdx$ by the generated configuration.
\end{itemize}

\subsubsection{Objective}
The pricing problem minimizes the reduced cost of a candidate
configuration. The reduced cost is expressed as:
\begin{equation}
\overline{c}_{\DemandIdx}
\! = \!
-\DualIdx_{\DemandIdx}^{\eqref{eq:onePathperDemand}}
+
\BWD_{\DemandIdx}
\sum_{\LinkIdx \in \LinkSett}
\DualIdx_{\LinkIdx}^{\eqref{eq:linkCapa}}
\delta_{\LinkIdx} +
\DualIdx_{\TimeIdx}^{\eqref{eq:budget}} \! \! \! \sum_{\SegmentKnIdx \in \SegmentKtSet} \!
 \left|\varphi_{\SegmentKnIdx} - \varphi_{\SegmentKnIdx}^{\Bar{\ConfigIdx}_{\DemandIdx, \TimeIdx-1}}
  \right|,
\label{eq:reduced_cost}
\end{equation}
which is linearized as:
\begin{alignat}{2}
\overline{c}_{\DemandIdx}
&=
-\DualIdx_{\DemandIdx}^{\eqref{eq:onePathperDemand}}
+
\BWD_{\DemandIdx}
\sum_{\LinkIdx \in \LinkSett}
\DualIdx_{\LinkIdx}^{\eqref{eq:linkCapa}}
\delta_{\LinkIdx} \nonumber\\
&+
\DualIdx_{\TimeIdx}^{\eqref{eq:budget}} \sum_{\SegmentKnIdx \in \SegmentKtSet} \left( \varphi_{\SegmentKnIdx}^{\Bar{\ConfigIdx}_{\DemandIdx, \TimeIdx-1}} +
  \left(1- 2\varphi_{\SegmentKnIdx}^{\Bar{\ConfigIdx}_{\DemandIdx, \TimeIdx-1}}
  \right)\varphi_{\SegmentKnIdx} \right).
\label{eq:reduced_cost}
\end{alignat}
A negative value of
$\overline{c}_{\DemandIdx}$ indicates that the generated configuration can
improve the current restricted master solution and should therefore be
added to $\ConfigSet_{\DemandIdx,\TimeIdx}$.

\begin{figure}[htb]
    \centering
    \includegraphics[width=\columnwidth]{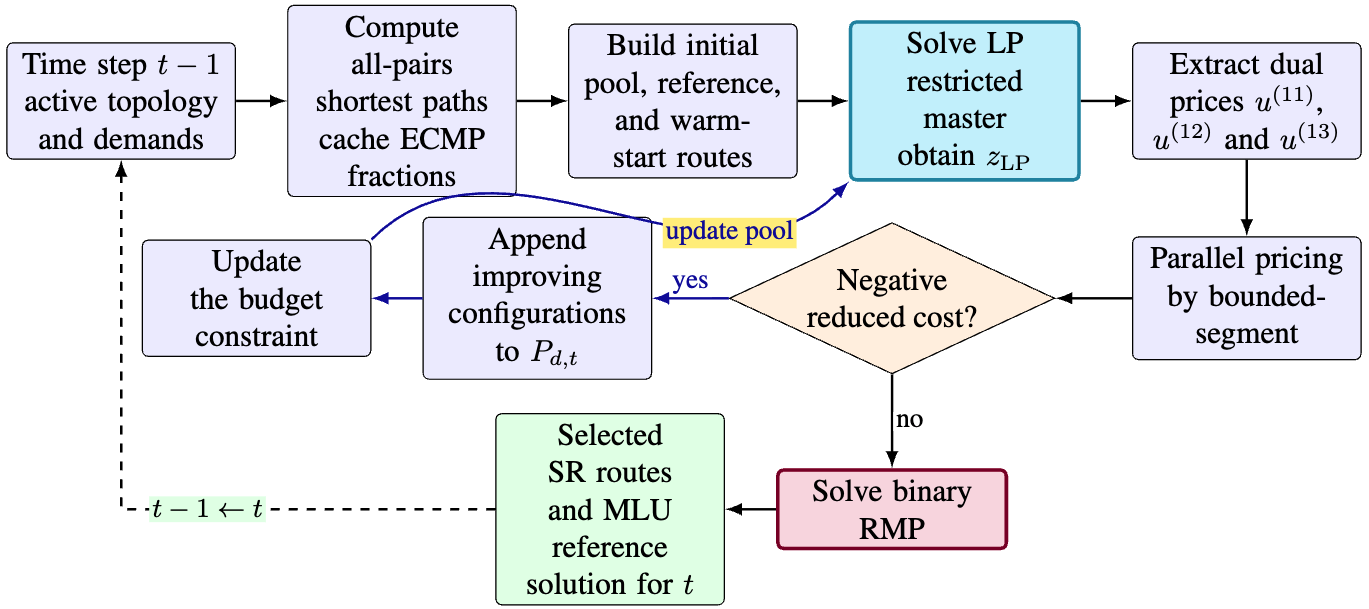}
    \caption{ILP Column-generation workflow: The RMP LP and pricing are iterated alternately until no improving configuration remains for the RMP LP, followed by an ILP solution of the last RMP.}
    \label{fig:cg_work_flow}
\end{figure}

\subsubsection{Constraints}

\paragraph{Flow conservation in the segment graph.}
The selected segments must form a path from the source to the destination:
\begin{align}
\sum_{\SegmentKnIdx \in \omega^+(v)}
\varphi_{\SegmentKnIdx}
-
\sum_{\SegmentKnIdx \in \omega^-(v)}
\varphi_{\SegmentKnIdx}
&=0,
&& v \in V\setminus
\{\SRC_{\DemandIdx},\DST_{\DemandIdx}\},
\label{eq:flow_conservation_segment}
\\
\sum_{\SegmentKnIdx \in \omega^+(\SRC_{\DemandIdx})}
\varphi_{\SegmentKnIdx}
&=1,
\label{eq:source_segment}
\\
\sum_{\SegmentKnIdx \in \omega^-(\DST_{\DemandIdx})}
\varphi_{\SegmentKnIdx}
&=1.
\label{eq:destination_segment}
\end{align}

Here, $\omega^+(v)$ and $\omega^-(v)$ denote, respectively, the sets of
outgoing and incoming segments of node $v$ in $K_{\TimeIdx}$.
\paragraph{Maximum number of segments.}
The generated configuration is limited to at most $\maxseg$ segments:
\begin{align}
\sum_{\SegmentKnIdx \in \SegmentKtSet}
\varphi_{\SegmentKnIdx}
&\leq \maxseg.
\label{eq:max_segments}
\end{align}

\paragraph{Mapping to physical links.}
The physical-link contribution of the generated configuration is obtained
by aggregating the forwarding contribution of its selected segments:
\begin{align}
\delta_{\LinkIdx}
&=
\sum_{\SegmentKnIdx \in \SegmentKtSet}
f_{\LinkIdx\SegmentKnIdx}
\varphi_{\SegmentKnIdx},
&& \LinkIdx \in \LinkSett.
\label{eq:mapping_segment_physical}
\end{align}

\paragraph{Variable domains.}
The decision variables satisfy
\begin{align}
\varphi_{\SegmentKnIdx}
&\in \{0,1\},
&& \SegmentKnIdx \in \SegmentKtSet,
\label{eq:phi_domain}
\\
\delta_{\LinkIdx}
&\geq 0,
&& \LinkIdx \in \LinkSett.
\label{eq:delta_domain}
\end{align}


\subsection{Column Generation and $\varepsilon$-optimal Integer Solution}

The pricing problem is solved for each demand using the current dual 
solution of the master problem. If no demand admits a configuration with
negative reduced cost, the solution of the continuous relaxation of the current restricted master problem is optimal
with respect to the complete configuration space. Denote by $\zLPOPTt$ (or $\zLB$ for short to indicate that $\zLPOPTt$ is a lower bound)  the corresponding value.

Otherwise, the generated
configurations are added to the master problem and the process is repeated
until no improving configuration remains. 

Once $\zLPOPTt$ has been derived, we solve the last remaining restricted master problem with an ILP solver. This does not guarantee an optimal ILP solution, but an upper bound ($\zILPUBt$ or $z_{\text{best}}$ for short, meaning the best known value) on the optimal ILP value ($\zILPOPTt$).

The flowchart depicted in Figure \ref{fig:cg_work_flow} summarizes the solution process.


\section{Results}
\label{sec:results}

\subsection{Data Sets}

We consider all Set A instances and a subset of the Set B instances from the 2026 Orange competition benchmark for time-varying segment routing, as the available computational resources did not permit us to solve the largest instances.
Each instance consists of three input files: a directed network, a time-indexed traffic matrix, and a scenario.
First, the network file specifies the Interior Gateway Protocol (IGP) metrics and positive link capacities.
For every directed link from node $u$ to node $v$, the reverse link from $v$ to $u$ also exists, and there is at most one link in each direction between any pair of nodes.
Second, the traffic matrix contains demands with pairwise distinct source-destination pairs $(\SRC_\DemandIdx,\DST_\DemandIdx)$; that is, no two demands share the same ordered pair. A traffic volume $\BWD_\DemandIdx$ is specified for each demand at each time slot.
Finally, the scenario file defines the maximum number of segments, the transition budgets, and the planned link removals.

Table~\ref{tab:dataset-characteristics} summarizes the main characteristics of Sets A and B.
On average, the Set B topologies contain 3.8 times as many nodes and 2.9 times as many links as Set A, while their demand matrices contain 5.6 times as many
traffic demands.

\begin{table}[htbp]
  \caption{Characteristics of the benchmark datasets.}
  \label{tab:dataset-characteristics}
  \centering
  \resizebox{\columnwidth}{!}{%
    \begin{tabular}{lcc}
      \hline
      \textbf{Characteristic} & \textbf{Set A} & \textbf{Set B} \\
      \hline
      Number of instances & 20 & 12 \\
      Nodes & 20--400 & 264--1263 \\
      Directed links & 80--2000 & 864--5036 \\
      Demands & 20--6000 & 2546--15000 \\
      Time slots & 2 & 12 \\
      Maximum segments per route & 6 & 8 \\
      Budgeted transitions & 1 & 11 \\
      Reconfiguration budget & 1--90 & 10--78 \\
      Planned link removals & One at $t=1$ & 2--10 per active slot \\
      \hline
    \end{tabular}%
  }
\end{table}

\subsection{Preliminary Results and Discussion}

\begin{figure*}[t]
  \centering
  \includegraphics[width=0.95\textwidth]{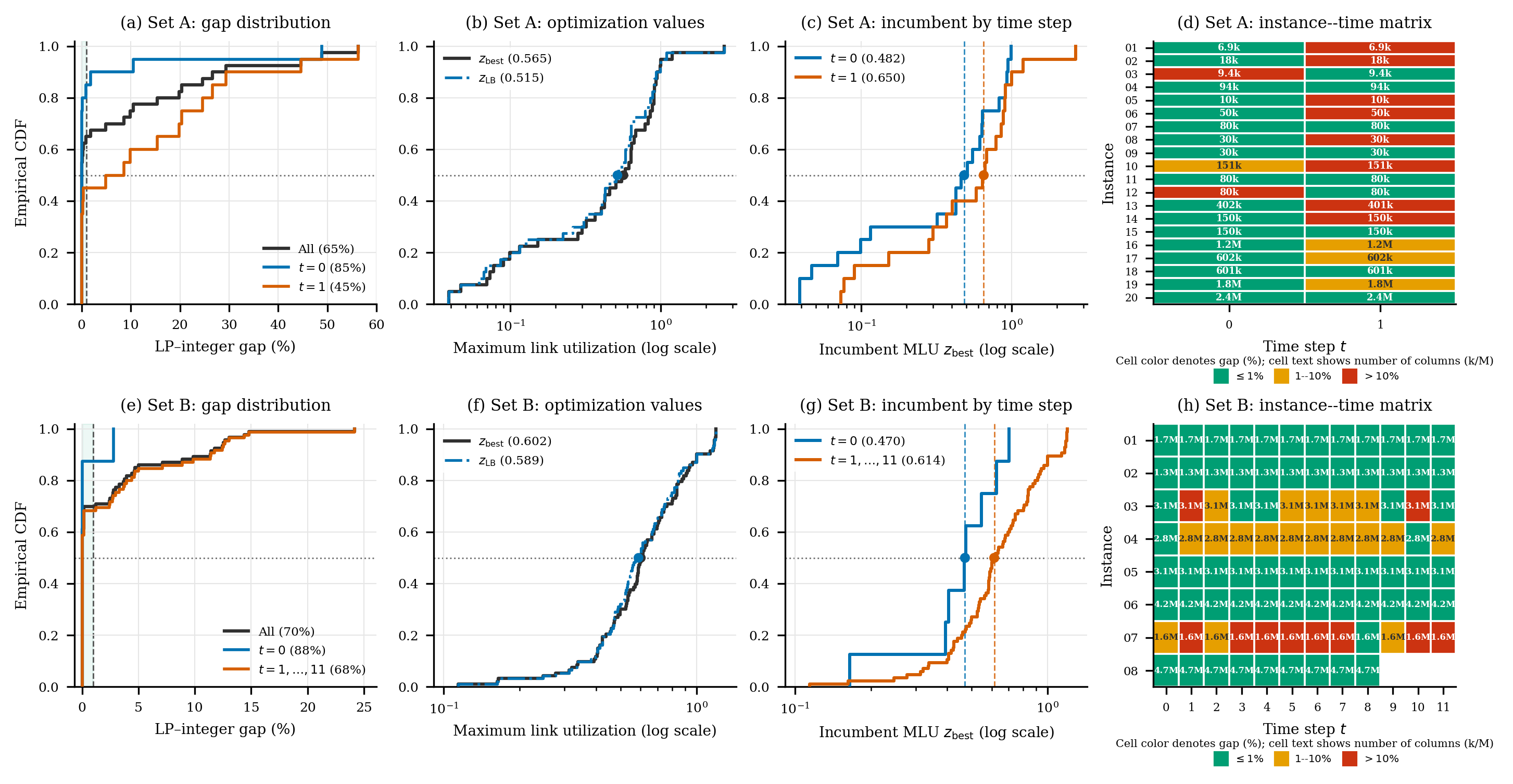}
  \caption{Solution-quality summaries for Set A (top row) and Set B (bottom row).
    Panels (a) and (e) show the LP-integer-gap CDFs, with shading for gaps no greater than 1\%.
    Panels (b) and (f) compare $z_{\mathrm{LP}}$ and $z_{\mathrm{best}}$, while panels (c) and (g) compare the incumbent MLU before and after the initial time step.
  Panels (d) and (h) show one cell for each instance-time-step pair: cell color indicates the LP-integer gap, while the annotation reports the rounded number of generated columns in compact k/M notation.}
  \label{fig:combined-cdfs}
\end{figure*}

Figure~\ref{fig:combined-cdfs} reports the accuracy of the solution output by the column-generation method for both data sets, and Table~\ref{tab:dataset-characteristics} summarizes the scale and temporal structure of the benchmark dataset.
Every run represented in this figure returned an integer solution, which is not guaranteed to be optimal.
The final {\NameMILP} lower bound matches the solution (an upper bound) to the reported accuracy in 87.97\% of the evaluations ($117/133$), and the largest remaining absolute bound-incumbent difference is $5.005\times10^{-3}$.
Within the generated configuration pool, the {\NameMILP} model was therefore solved exactly in most runs, while the remaining runs retained small but nonzero optimality gaps.
We measure the accuracy of the ILP solution provided by the {\NameMILP} algorithm as follows
\begin{equation}
  \varepsilon_{t}(\%)
  = 100\,
  \frac {\zbestt - \zLBt} {\zLBt},
  \label{eq:lp-integer-gap}
\end{equation}
where $\zLBt$ is the optimal MLU of the LP-relaxation of the master problem at time $t$.
Therein, $0 \leq y_{\DemandIdx}^{\ConfigIdx} \leq 1$, so the LP solution may combine several configurations for the same demand.
The value $\zbestt$ is the MLU of the best feasible integer solution, for which $y_{\DemandIdx}^{\ConfigIdx} \in \{0,1\}$ and exactly one configuration is selected for each demand.
A zero reported gap means that $\zbestt = \zLBt$, i.e., $\zbestt$ is an optimal ILP solution.
A larger gap indicates looser bounds, either lower or upper one, or both.

For Set A, Fig.~\ref{fig:combined-cdfs}(a) shows a right-skewed gap distribution: 
26 of 40 evaluations (65\%) have gaps no larger than 1\%, including 18 (45\%) with zero gaps.
Although the median is only 0.015\%, the mean reaches 8.09\% because 25\% of the evaluations have gaps above 10\% and 17.50\% exceed 20\%.
The gaps are more pronounced after the network transition.
At $t=0$, 85\% of the evaluations are within 1\%, the median is zero, and the mean is 3.12\%; 
at $t=1$, these values change to 45\%, 6.75\%, and 13.05\%, respectively.
Figure~\ref{fig:combined-cdfs}(b) shows that $\zLB$ and $\zbest$ coincide in 97.50\% of the evaluations, 
indicating that the large LP-integer gaps generally do not require the use of a branch-and-price algorithm (i.e., an exact algorithm for solving the master problem with integrality requirements) to reach an optimal ILP solution.
Figure~\ref{fig:combined-cdfs}(c) also shows a rightward shift of the incumbent distribution after the transition, when routing decisions must avoid unavailable links while remaining close to the previous configuration.
The LP relaxation can respond by combining several configurations for one demand, whereas the integer model must select one complete configuration.
This difference can widen the relaxation gap when the transition budget is binding.

The expanded Set B sample shows a more heterogeneous pattern despite the larger networks described in Table~\ref{tab:dataset-characteristics}.
As shown in Fig.~\ref{fig:combined-cdfs}(e), 70\% of the 93 evaluations have gaps no greater than 1\%, 52.69\% have zero gaps, and the median remains zero; the mean is 2.33\%, and the maximum is 24.19\%.
At $t=0$, 87.50\% of the evaluations are within 1\%, and the mean gap is 0.35\%; over the subsequent periods, these values become 68.24\% and 2.51\%, respectively.
The median incumbent MLU also increases from 0.470 at $t=0$ to 0.614 after the initial period, consistent with the combined effect of link removals and the requirement to remain close to the preceding routing configuration.

The additional instances show that weak bounds are not confined to one trajectory.
As shown in Fig.~\ref{fig:combined-cdfs}(h), gaps above 1\% occur in seven setB-03 time periods, ten setB-04 time periods, and eleven setB-07 time periods.
Ten evaluations (10.75\%) exceed 10\%: two belong to setB-03 and eight to setB-07.
The temporal profiles are nevertheless different.
setB-03 has isolated spikes, including the overall maximum at $t=10$; setB-04 has moderate but recurring post-transition gaps; and setB-07 has a persistent relaxation gap with a median of 11.58\%.
The column-pool results in Fig.~\ref{fig:combined-cdfs}(d) and (h) provide a separate view of solution quality.
For Set A, the pool ranges from approximately $6.9\times10^{3}$ to $2.4\times10^{6}$ configurations, but larger pools do not consistently correspond to either larger or smaller gaps.
The expanded Set B pools range from approximately $1.3$ to $4.7$ million configurations, and the pooled rank correlation between column count and the LP-integer gap is weak and slightly negative (Spearman $\rho=-0.14$).
setB-07 illustrates the disconnect: its pools contain about $1.57$ million configurations but its median gap is 11.58\%, whereas setB-05, setB-06, and setB-08 use roughly $3.1$-$4.7$ million columns and have gaps no greater than 0.17\% in the available periods.
Together, these results indicate that column count tracks instance scale more closely than relaxation strength.
Therefore, generating more columns indiscriminately is unlikely to resolve the difficult transition cases.
More targeted improvements include 
stabilization techniques for the dual prices, and generating in advance good alternate routing for demands using congested or removed (under maintenance) links during a given period.

Across both sets, 68.42\% of the 133 evaluations have an LP-integer gap no greater than 1\%, and 50.38\% have a zero gap.
Thus, the generated column pools often yield tight fractional relaxations, but the setB-03, setB-04, and setB-07 trajectories expose clear instance-dependent and transition-dependent limitations.
Accordingly, a global certificate for the complete configuration space may require a branch-and-price algorithm, or some of the steps of it in order to reach proven optimal integer solutions.
The difficult transition cases motivate further work on reconfiguration-aware pricing problems and the integrality constraints.


\section{Conclusions}

We investigated segment routing under the classical MLU objective and subject to reconfiguration budget constraints when reconfiguration is required in order to take care of, e.g., some link maintenance. 
Budget constraints can be rewritten in order to become linear constraints, and consequently the proposed column generation model is able to solve pretty large data instances and to output solutions with a very good accuracy. 

Future work will include improvement of the current model with different decomposition schemes, as well as identification of some  patterns of the demand routing when it is not associated with the shortest path(s). This will speed up the resolution process, in accordance with the principle that optimization techniques excel at exploiting optimality patterns to achieve good results.

\bibliographystyle{IEEEtran}
\bibliography{Biblio/biblio,Biblio/BJ,Biblio/LP}

\end{document}